\documentclass[pra,twocolumn,showpacs]{revtex4}

\usepackage{bm,amsmath,amsfonts,amsbsy}

\def \d3k {\frac{d^3k}{(2\pi)^3}}
\def \v#1{{\bm #1}}
\def \in {{\rm in}}
\def \be {\begin{equation}}
\def \ee {\end{equation}}

\begin{document}

\title{Radiation from accelerated impurities in Bose-Einstein condensate}
\author{Jun Suzuki}
%\email{junsuzuki@nii.ac.jp}
%\author{Jun Suzuki\footnote{Present address: National Institute of Informatics, 
%2-1-2 Hitotsubashi, Chiyoda-ku, Tokyo 101-8430, Japan}}
\address{National Institute of Informatics, 2-1-2 Hitotsubashi, Chiyoda-ku, Tokyo 101-8430, Japan}
%\address{Department of Physics and Astronomy, University of South Carolina, Columbia, SC 29208}
\date{\today}

\begin{abstract}
We investigate radiation spectra arising from accelerated point-like impurities in the homogeneous 
Bose-Einstein condensate. A general formula for the radiation spectrum is obtained in the integral form as a function 
of given impurity trajectory. The Planckian spectrum is obtained for a special accelerated motion, which is shown 
to be unphysical. Non-Planckian spectrum is found in the case of a uniformly accelerated impurity. We 
compare our result with similar settings as discussed in other quantum many-body systems. 
\end{abstract}

%\begin{keyword}
%Bose-Einstein condensation \sep Accelerated impurities \sep Energy dissipation
%\PACS 03.75.Kk \sep 03.70.+k \sep 67.40.Yv
%\end{keyword}

\maketitle

 %%%%%%%%%%%%%%%%%%%%%%%%%%%%%%%%%%%%
\section{Introduction}

It is now a well-established experimental fact that condensation of mesoscopic 
number of atoms can be created and manipulated for relatively long time-scale on the order of seconds. 
These experimental progress enable us to probe properties of ultra-cold 
gases and Bose-Einstein condensate (BEC) from various different perspectives. 
Among them, many authors have been investigating the response properties of BECs 
when time-dependent perturbations are applied \cite{meystre,kovrizhin,astracharchik,mazets,timmermans,jun1,jun2}. 
On general grounds, time-dependent disturbance on quantum many-body system will give rise to 
energy dissipation and hence to destroy the order and coherence in the system after some time later. 
However, due to the phenomenon of superfluidity of BECs, energy dissipation 
may or may not take place when time-dependent perturbation is applied to the condensates. 
An important question is to know how and when such a dissipation process occurs. 
This paper studies the radiation spectra accompanied by accelerated motions of an impurity 
in the homogeneous condensate. The motions of classical impurity 
is a special class of time-dependent disturbance and can be realized by using a detuned laser, 
a small number of different species of atoms, or many other microscopic objects in BECs. 
We are interested in the form of spectra obtained for various kinds of accelerated motions of the impurity. 

Recently, the idea of simulating relativistic and cosmological effects in non-relativistic 
quantum (many-body) systems has gained great attention from various fields of researchers. 
There have been many activities devoted to explore this idea and how to realize it in laboratories. 
Several authors have studied radiation spectra arising from accelerated 
disturbance and responses of accelerated detectors in several different quantum many-body systems 
\cite{SKBFC,ADM,RCPR,MBM,laserUnruh}. 
They found the Planckian distribution for the radiation spectra 
and for the detector's excitation probabilities. These results were then interpreted with a simulation of 
the Fulling-Davies-Unruh (FDU) effect in which {\it any accelerated observer} perceives the empty 
Minkowski vacuum as a finite temperature equilibrium state \cite{fulling,davies,unruh}. 
To be more precise, these studies concern simulating the DeWitt-Unruh (DU) effect and the FDU radiation. 
The former states that {\it a particular accelerated detector} is excited in the same way as being replaced 
in a thermal bath at rest and the latter concerns the phenomenon of thermal radiation from an accelerated 
particle in the vacuum \cite{takagi,CHM}. Regarding the DU effect, there exist several 
non-thermal response for the detector \cite{NR,MKP,FNMB}. 
There also exists a debate if the original derivation of the FDU effect is mathematically correct 
or not \cite{belinskii}. It is questioned in this paper whether or not such an appearance of Planckian distribution 
in an analog situation can be related to the FDU effect. We briefly discuss the consequence 
of our result for proposed emergent theories of gravity in quantum many-body systems. 

In this paper we investigate a coupling of a classical point-like impurity to the homogeneous condensate 
at zero temperature and study the radiation spectra accompanied 
by the accelerated motions of impurity. 
The impurity is assumed to move along a given trajectory without any 
back reactions. When the total number of bosons in the system is very large, the effect of 
point-like impurity can be represented by the time-dependent driven oscillator model, which 
can be diagonalized by a suitable unitary transformation. The radiation spectrum 
is obtained in the integral form as a function of a given impurity trajectory. 
In this paper, the Planckian spectrum is also obtained when the impurity is accelerated 
in a special manner inside the homogeneous condensate. However, this special class of impurity trajectory 
is shown to be unphysical and hence would not be realized to simulate such a black-body radiation. 
A radiation spectrum is also obtained analytically for a uniform acceleration and they are analyzed in detail. 

The outline of this paper is as follows. Sec.~II provides a model and its diagonalization 
within the Bogoliubov approximation. We obtain the radiation spectra for given impurity 
trajectories in Sec.~III. Two accelerated motions are studied and are analyzed with their corresponding 
radiation spectra in Sec.~IV. We discuss our result and its connection to the FDU effect in Sec.~V. 

%%%%%%%%%%%%%%%%%%%%%%%%%%%%%%%
\section{Model Hamiltonian}
%%%%%%%%%%%%%%%%%%%%%%%%%%%%%%%

The model Hamiltonian for weakly interacting bosons of mass $M$ is 
\begin{multline} \label{h0}
{H}_0 = \int\!d^3x\;{\psi}^{\dagger} (x) \left(-\frac{\hbar ^2 \v{ \nabla} ^2}{2 M}\right){\psi} (x)\\
+ \frac g2  \int\! d^3x\; {\psi}^{\dagger} (x){\psi}^{\dagger}(x){\psi} (x){\psi} (x) ,
\end{multline}
where the general two-body interaction term between bosons is approximated 
by a contact interaction $g \delta(\v x-\v x')$ in the low energy region. 
The field operators ${\psi}^{\dagger}(x)={\psi}^{\dagger} (\v{ x},t)$ 
and ${\psi} (x)={\psi} (\v{ x},t)$ in the Heisenberg picture 
satisfy the equal time canonical commutation relations. 
We assume that the repulsive interaction $g>0$ and the condition of diluteness $na_s^3\ll1$, 
where $n$ is the number density of massive bosons. 
We consider the following local interaction Hamiltonian between bosons and a point-like classical impurity 
moving along a given trajectory $\v{ \zeta}(t)$: 
\be \label{hi}
{H}_{{\rm I}} (t) = \lambda  \int d^3x\; \delta (\v x -\v{ \zeta}(t)){\psi}^{\dagger} (x){\psi} (x) , 
\ee
where $\lambda$ represents the coupling constant between bosons and the impurity. 
As we discuss later, $\lambda$ can be time-dependent in general 
when the impurity coupling is switched on and off in experiments. 
For the homogenous bose gas under consideration, the field operators are expanded 
in terms of the plane wave basis with 
periodic boundary conditions in a finite size box $V=L^3$ to obtain the total Hamiltonian 
${H}={H}_0+{H}_{{\rm I}}$ as 
\begin{multline} \label{h1}
{H}= \sum _{\v k} \epsilon _{k}{a}_{\v k}^{\dagger}{a}_{\v k}
+ \frac{g}{2V} \sum _{\v k,\v k',\v q}{a}_{\v k+\v q}^{\dagger}{a}_{\v k'-\v q}^{\dagger}{a}_{\v k}{a}_{\v k'}\\
+ \frac{\lambda}{V} \sum _{\v k,\v k'} e^{-i\v k \cdot \v{ \zeta}(t) }{a}_{\v k+\v k '}^{\dagger}{a}_{\v k}, 
\end{multline}
where $ \epsilon _{k}=\hbar ^2 \v k^2/2M$ is the free kinetic energy of bosons 
and the summation is taken over integers $n_i$ for $\v k=2\pi(n_x,n_y,n_z)/L$. 
In the following we are interested in the effects of impurities to the homogeneous condensate 
in the large $N$ limit. We will investigate the first order correction on the homogeneous condensate. 

We follow Bogoliubov's treatment to simplify the full Hamiltonian within the number-fixed framework \cite{NCmethod}. 
Define the operator $ {\beta}_0 = ({N}_0 +1)^{-1/2}{a}_0 $ where ${N}_0 = {a} ^{\dagger}_{0}{a}_{0} $ 
is the number operator of condensed bosons. This operator and its hermite conjugate satisfy 
${\beta} _{0}{\beta}^{\dagger}_{0}=1$ and ${\beta}^{\dagger}_{0}{\beta}_{0}=1-{\Pi}_0$, 
where ${\Pi}_0=|N_0=0\rangle\langle N_0=0|$ is the projection operator onto the state $N_0 =0$. 
In the presence of the homogeneous BEC for the zero mode, we can exclude the state $N_0 =0$ 
to approximate ${\beta}^{\dagger}_{0}{\beta}_{0} \simeq 1$, 
i.e., ${\beta}_0$ is an almost unitary operator, and $[ {\beta} _{0}, {\beta}^{\dagger}_{0}]\simeq 0 $. 
More precisely this approximation holds between matrix elements when we compute expectation values. 
Then ${a}^{\dagger}_0=\sqrt{{N}_0} {\beta}^{\dagger}_0$ and ${a}_0= {\beta}_0 \sqrt{{N}_0} $ 
allow us to eliminate the bare zero mode operators ${a}^{\dagger}_0$ and ${a}_0$. 
We then introduce a new set of the creation and the annihilation operators 
${\alpha}^{\dagger}_{\v k}$ and ${\alpha}_{\v k}$ for non-zero momentum: 
$ {\alpha}^{\dagger}_{\v k} = {a}^{\dagger} _{\v k}  {\beta}_0 , \ {\alpha}_{\v k} ={\beta}^{\dagger}_0 
{a} _{\v k} \quad {\rm for}\  \v k\neq 0 $. They satisfy the equal time canonical commutation relations: 
$ [{\alpha}_{\v k}(t), {\alpha} ^{\dagger}_{\v k'}(t)]  = \delta_{\v k,\v k'}- {\alpha}_{\v k'} 
{\alpha} ^{\dagger}_{\v k} {\Pi}_0 \simeq \delta_{\v k,\v k'}  ,\quad [{\alpha}_{\v k}(t), {\alpha} _{\v k'}(t)] 
 =[ {\alpha} ^{\dagger}_{\v k}(t), {\alpha} ^{\dagger}_{\v k'}(t)]=0 $. 
The vacuum state of the Fock space ${\cal H}_0$ for the new set of the creation and the annihilation operators 
is the same as the plane wave basis, i.e., ${\alpha}_{\v k} |0 \rangle =0$. 

The total number operator ${N}={N}_0+\sideset{}{'}\sum  {\alpha}_{\v k}^{\dagger}  {\alpha}_{\v k}$
commutes with the total Hamiltonian, where the prime over the summation is used to indicate 
the omission of the zero mode for the summation. Therefore we can choose the basis in which ${N}$ 
is diagonalized with given value $N$. This gives the condition 
${N}_0+\sideset{}{'}\sum  {\alpha}_{\v k}^{\dagger}  {\alpha}_{\v k}=N$ 
which allows us to eliminate the zero mode occupation number operator ${N}_0$. 
Using an expansion for the square root function we can proceed to do a systematic expansion in $1/\sqrt{N}$. 
The Bogoliubov transformation to diagonalize the Hamiltonian without an impurity term is $U_B=\exp (i {G}_B)$; 
\begin{align} \label{GB}
{G}_B &= \frac i2 \sideset{}{'}\sum \theta _{k}  {\alpha}_{\v k}^{\dagger}  {\alpha}_{-\v k}^{\dagger}+{\rm h.c.},\\
 \theta _{k} &= \tanh ^{-1}(\frac{gn}{\hbar \omega _{k} +\epsilon _{k}+gn}), \\
 \hbar \omega _{k}&= \sqrt{\epsilon _{k}(\epsilon _{k}+2 gn)} .
\end{align}
Bogoliubov's excitation (the ``bogolon") is created and annihilated by the operators
\begin{align} \label{bog1}
{b}_{\v k}^{\dagger}  &=
U_B{\alpha}_{\v k} ^{\dagger}U_B^{\dagger}
= {\alpha}_{\v k} ^{\dagger} \cosh \theta _{k} +{\alpha}_{-\v k} \sinh \theta _{k}, \\
{b}_{\v k}& = U_B {\alpha}_{\v k} U_B^{\dagger}
={\alpha}_{\v k}  \cosh \theta _{k} + {\alpha}_{-\v k}^{\dagger} \sinh \theta _{k},\label{bog2}
\end{align} 
respectively. Keeping up to the second order in the new set of creation and annihilation operators, 
we obtain the effective Hamiltonian for the impurity motion in the homogeneous BEC as 
\be \label{hb}
{H}= E_0
+\sideset{}{'} \sum \hbar \omega _{k} {b}_{\v k}^{\dagger}  {b}_{\v k}
+ \sideset{}{'} \sum \v (f_{\v k}(t) {b}_{\v k}^{\dagger}+{\rm h.c.}\v ) .
\ee
Here $E_0= g n N/2 -gn/2+ \lambda n + \sideset{}{'} \sum  (\hbar \omega _{k} -\epsilon _{k}-gn)/2$
is the ground state energy without impurity and
\be \label{f}
f_{\v k}(t)= n \lambda \sqrt{\frac{\epsilon _{k}}{N \hbar \omega _{k}}}\; e^{-i\v k \cdot \v{ \zeta}(t)} .
\ee
The homogeneous condensate is defined by the new vacuum state of the Fock space ${\cal H}_B$: 
${b}_{\v k} | BEC \rangle =0$.  

Thus, the impurity effects in our model are described by a time-dependent 
driven oscillator model, or the time-dependent van Hove model. It is noted that the effective interaction 
Hamiltonian, the third term in eq.~\eqref{hb}, is changed from the density coupling \eqref{hi} 
to the linear coupling; $\sqrt{n}\lambda\int dx^3\delta (\v x -\v{ \zeta}(t))({\tilde{\psi}}^{\dagger}+ {\tilde{\psi}})$, 
where ${\tilde{\psi}}$ is the bosonic filed operator excluding zero mode. 
Note that the factor $1/\sqrt{N}$ in Eq. \eqref{f} is due to the condensation which 
introduces the significant difference and gives rise to the macroscopic effect. 

We can solve the above Hamiltonian \eqref{hb} exactly with the initial condition 
such that the operators evolve without impurity asymptotically at the infinitely remote past $t=- \infty$. 
The solution is
\begin{align} \label{solution1}
{b}^{\dagger}_{\v k}(t)&={b}^{\in\dagger}_{\v k}(t)+\phi ^*_{\v k}(t),\\
{b}_{\v k}(t)&={b}^{\in}_{\v k}(t)+\phi _{\v k}(t) ,\label{solution2}
\end{align}
where $\phi _{\v k}(t)$ is a $c$-number function 
\begin{align}  \label{phi}
\phi _{\v k}(t) &= - \frac{in\lambda}{\hbar} \sqrt{\frac{\epsilon _{k}}{N \hbar \omega _{k}}}I_{\v k} (t)e^{-i \omega _{k} t}  ,\\ 
\label{integral}
I_{\v k}(t) &=\int ^t _{-\infty} \!d t'\;e^{i\omega_k t'-i\v k \cdot \v{ \zeta}(t')}.
\end{align} 
When the impurity is injected at some time $t_i$, the solution is given 
instead by the similar formulae (\ref{solution1},\ref{solution2},\ref{phi},\ref{integral}) with the replacement 
of the time integral $\int ^t _{-\infty}\to\int ^t _{t_i}$. 

The asymptotic creation and annihilation operators denoted by ${b}^{\in\dagger}_{\v k}$ and 
${b}^{\in}_{\v k}$ are governed by the Heisenberg equation with the diagonalized Hamiltonian 
${H}_{\in}= \tilde{E}_0(t)+ \sideset{}{'} \sum \hbar \omega _{k} {b}^{\in \dagger}_{\v k} {b}^{\in}_{\v k}$. 
In the diagonalized Hamiltonian, the new ground state energy is defined by 
$\tilde{E}_0(t)= E_0+ \sideset{}{'} \sum {\rm Re}\v (f_{\v k}^*(t)\phi _{\v k}(t)\v )$. 
The energy spectrum $\hbar \omega _{k}$ is that of the gapless excitations characterized 
by $\omega _{k} \simeq kc$ for a small $k$, where $c=\sqrt{gn/M}$ is the speed of sound of the condensate. 
This is when $k\xi\ll 1$ holds, where $\xi=\hbar/(2Mc)$ is the coherence length of the condensate. 
For large $k$, on the other hand, $\hbar \omega _{k}\simeq\epsilon_k$ which is the free 
particle energy spectrum.  
We remark that the spectrum $\omega_k$ and the speed of sound $c$ are the same 
as in the original Bogoliubov model without impurities. 
Therefore the motions of impurity do not affect neither $\omega_k$ nor $c$ within our approximation.

The vacuum state of the Fock space $\cal{H}_{\in}$ for ${b}^{\in\dagger}_{\v k}$ and
${b}^{\in }_{\v k}$ is defined by ${b}^{\in }_{\v k}\ |0_{\in}\rangle =0$. 
%The relation between two vacua of Fock spaces ${\cal H}_{B}$ and ${\cal H}_{\in}$ is formally expressed as
%\be
% |0_{\in} \rangle = \exp [ \sideset{}{'}\sum \v (-\frac 12  | \phi _{\v k}(t) |^2 + \phi _{\v k}(t) {b}_{\v k}^{\dagger} \v ) ] | BEC \rangle .
%\ee
At remote past $t=-\infty$, the vacuum $ |0_{\in}\rangle$ coincides with the homogeneous condensate 
$|BEC\rangle$. The physical interpretation of the above result is quite simple. The motions of impurity 
creates the time-dependent coherent states. The dressed Bogoliubov's excitations do not annihilate 
the homogeneous condensate vacuum: 
\be \label{result}
{b}^{\in}_{\v k} \ |BEC\rangle =-  \phi _{\v k}(t) |BEC\rangle .
\ee

%%%%%%%%%%%%%%%%%%%%%%%%%%%%%%%
\section{Radiation spectrum}
%%%%%%%%%%%%%%%%%%%%%%%%%%%%%%%

One of main effects described by our model is the effect of energy radiation 
due to the motions of impurity in the homogeneous condensate. From our model we find 
that even though the systems is in the ground state where no elementary excitation 
is present, there is energy transferred from the motions of impurity to massive bosons. 
In other words, the motions of impurity will dress bogolons in such a way that bogolons 
will not see the homogeneous condensate as the vacuum. Since we do not take into account 
any back reaction on the impurities, this transferred energy is the energy required in order for 
the impurities to move along given trajectories. It means that we need to feed this amount 
of energy to keep the impurities moving. More realistic model for the radiation effects in BEC 
should be done by considering quantum impurities and taking into account the dynamics of them.  

The occupation number for bogolons with respect to the homogeneous condensate 
\be \label{occu}
\tilde{n}_{\v k}(t) \equiv \langle BEC|\ {b}^{\in\dagger}_{\v k} {b}^{\in}_{\v k} \ |BEC \rangle =\frac{n^2 \lambda ^2 \epsilon_k}{N \hbar^3 \omega_{k}}|I_{\v k}(t)|^2, 
\ee
counting the emitted bogolons accompanying with the motions of impurity in BEC. 

We next evaluate the expectation value for the number operator 
${N}_{\v k} (t)= {a}_{\v k}^{\dagger}(t){a}_{\v k}(t)={\alpha}_{\v k}^{\dagger}(t){\alpha}_{\v k}(t)$ 
for the excited particles with respect to the vacuum $\langle 0_{\in}| {N}_{\v k} (t) |0_{\in} \rangle$. 
This number counts the number of non-condensated particles. 
The depletion of the condensate $d(t)$ due to the quantum fluctuation 
can be evaluated by summing over all modes and divided by the total particle number $N$:
\begin{align}
d(t)&=\frac1N \sideset{}{'} \sum \langle 0_{\in}| {N}_{\v k} (t) |0_{\in} \rangle \\ \nonumber
&= \frac 83 \sqrt{\frac{n a_s^3}{\pi}}+ \frac 1N  \sideset{}{'} \sum (\frac{\epsilon _{k}}{\hbar \omega _{k}}
|\phi _{\v k}(t)|^2\\
&\hspace{2cm}+\frac{gn}{2\hbar \omega _{k}}|\phi^*_{\v k}(t)-\phi_{-\v k}(t)|^2) .
\end{align}
Since $|\phi_{\v k}|^2$ has an additional factor $1/N$, the motions of impurity do not contribute 
to the depletion of the homogeneous condensate in the thermodynamic limit. In contrast, real experiments 
have always a finite number of particles and size. Therefore, there are effects on the depletion of 
condensates due to the motions of the impurity.

The dissipated energy ${\cal E}_{\v k} (t)$ for a given mode $\v k$ is obtained by multiplying 
$\tilde{n}_{\v k}(t)$ by the excitation energy $\hbar \omega _k$, i.e., 
${\cal E}_{\v k} (t)=\hbar \omega _k \tilde{n}_{\v k}(t)$ and the total dissipated energy ${\cal E} (t)$ 
is given by summing over all modes. 
Therefore, for the $t \to \infty$ limit which takes into account the complete trajectory of impurity motion, 
we obtain rather a simple expression the total dissipated energy in the thermodynamic limit as 
\begin{multline} \label{diss}
{\cal E}_{{\rm tot}}=\lim _{t \to \infty} \sideset{}{'} \sum {\cal E}_{\v k} (t)\\
\to \frac{n \lambda ^2 }{ \hbar^2} \int \d3k \epsilon_k \;
\;\left| \int_{-\infty}^{\infty}dt \;e^{i\omega_k t-i\v k \cdot \v{ \zeta}(t)}\right|^2.
\end{multline}
When an impurity is moving with a constant velocity, i.e., $\v{\zeta} (t)=\v{v} t$, 
the total dissipated energy per unit time is proportional to 
$ \int d^3k \epsilon_k \delta(\omega_k-\v k\cdot\v v)$. This clearly shows that 
the speed of impurity $v=|\v{v}|$ needs to exceed the speed of sound to 
create finite amount of radiation, otherwise there is no radiation from 
the impurity motion \cite{kovrizhin,astracharchik,jun1,jun2}. 
When the impurity acquires some acceleration, on the other hand, 
the integral \eqref{diss} produces a non-zero value, that is a finite amount of energy radiation. 

%%%%%%%%%%%%%%%%%%%%%%%%%%%%%%%
\section{Radiation from accelerated impurities}
%%%%%%%%%%%%%%%%%%%%%%%%%%%%%%%

We consider three examples to illustrate our model and its general solution. 
We are interested in the occupation number (\ref{occu}) and the total dissipated energy 
(\ref{diss}) for the $t\to \infty$ limit, i.e., in evaluating the integral: 
\be \label{int} 
I_{\v k} =\lim_{t\to\infty}I_{\v k} (t)=\int _{-\infty}^{\infty}dt \;e^{i\omega_k t-i\v k \cdot \v \zeta (t)}.  
\ee
This limit, of course, should be understood as an idealization to obtain 
analytical results, and one needs to calculate a finite interval to compare 
with experiments. In the following, we also calculate the numbers of emitted bogolons $d\tilde{n}_{\v k}$ 
between $\v k$ and $\v{k}+d\v{k}$ in the thermodynamic volume limit. 
Similarly, the energy radiated between $\v k$ and $\v{k}+d\v{k}$ is $ d{\cal E}_{k} =\hbar\omega_k d\tilde{n}_{\v k}$. 
We note that a constant translation to the impurity trajectory $\v \zeta (t)\to\v \zeta (t)+\v \zeta_0$ 
gives rise to a phase factor which vanishes upon taking the absolute square. 
Thus any translation of the impurity trajectory will not produce any physical effect. 

%%%%%%%%%%%%%%%%%%%%%%%%%%%%%%%
\subsection{Black-body-like radiation}
%%%%%%%%%%%%%%%%%%%%%%%%%%%%%%%
We first seek a trajectory which leads to the Planckian distribution for the 
occupation number \eqref{occu}. With a simple analysis, the trajectory with 
two parameters $\zeta_0$ and $\Gamma_0>0$ 
\be \label{ther}
\v \zeta (t)=\zeta_0 e^{-\Gamma_0 t}\; (0,0,1), 
\ee
leads to the integral 
\be \label{int1} 
I_{\v k} =\int _{-\infty}^{\infty}dt \;\exp(i\omega_k t-i k_z \zeta_0 e^{-\Gamma_0 t}).  
\ee
Here $k_z$ is the $z$ component of the wave number vector of the emitted bogolons. 
The parameter $\Gamma_0$ can be expressed in terms of the acceleration constant $a$ as $\Gamma_0=a/c$ 
and we refer it to as the acceleration parameter. The case of negative acceleration $\Gamma_0<0$ 
can be analyzed similarly and we only consider the positive case in this paper. 
Changing the variable from $t$ to $\eta=e^{-\Gamma_0 t}$ reads 
\be \label{int1-1} 
I_{\v k} =\frac{1}{\Gamma_0} \int _{0}^{\infty}d\eta \;\eta^{-i\omega_k /\Gamma_0-1} e^{-i k_z \zeta_0 \eta}.  
\ee
To carry out this integral, one formally introduces the regularization 
$\omega_k /\Gamma_0\to \omega_k /\Gamma_0+i\epsilon$ where 
$\epsilon$ is a small positive number which is set to zero at the end of calculations. 
Such a regularization, however, needs to be justified and this will be discussed in the next section. 
It is important to note that the $z$ component of the wave vector $k_z$ can be both positive or negative. 
With the above mentioned regularization, we obtain the integral as
\be\label{int1-2}
I_{\v k}=\frac{1}{\Gamma_0}\left|k_z\zeta_0\right|^{i \frac{\omega_k }{\Gamma_0}}
\Gamma \left(-i\frac{\omega_k }{\Gamma_0}\right)
e^{-{\rm sgn} (k_z\zeta_0)\; \frac{\pi\omega_k }{2 \Gamma_0}}, 
\ee
where ${\rm sgn}(x)$ is the sign function and $\Gamma(x)$ is the Gamma function. 

The appearance of the above integrals \eqref{int1} is rather surprising when 
comparing with an informal derivation of the FDU effect using time-dependent 
Doppler shift, see in particular eq.~(7) of Ref.~\cite{AM}. 
Similar formulae have been used by many authors to discuss FDU-like effects in non-relativistic systems 
\cite{SKBFC,ADM,RCPR}. 
However our calculation exhibits a different result as discussed in the next section. 
 
The occupation number is 
\begin{multline} \label{result1}
\tilde{n}_{\v k}=\frac{n^2 \lambda ^2 \epsilon_k}{N \hbar^3 \omega_{k}} \frac{2\pi }{\omega_k \Gamma_0}\\
\hspace{0.5cm}\times
\begin{cases}
\left[e^{2\pi \omega_k /\Gamma_0}-1\right]^{-1}\quad (k_z \zeta_0>0) \\
\left[1-e^{-2\pi \omega_k /\Gamma_0}\right]^{-1}\quad (k_z \zeta_0<0)
\end{cases}.
\end{multline}
When the direction of emitted bogolon is on the $xy$ plane, i.e., $k_z=0$, 
the above integral is proportional to the Dirac delta distribution $\delta(\omega_k)$. 
Using the formula $x\delta(x)=0$ together with the factor $\epsilon_k/\omega_{k}$, 
we get zero radiation for $k_z \zeta_0=0$. We now examine this result \eqref{result1} 
for the infrared and the ultraviolet region in detail.  
From now on, we choose the impurity trajectory to be confined within the positive $z$ space, 
i.e., $\zeta_0>0$,  and hence the impurity is accelerated from a positive infinity point $+\infty$ 
toward to the origin as the time changes from $-\infty$ to $+\infty$. First of all, we observe that 
the distribution of emitted bogolons does not depend on the direction within 
the upper half volume ($z>0$) and the lower half volume ($z<0$), respectively. Secondly, the radiation 
is independent of a scaling with respect to the parameter of the trajectory $\zeta_0$ and is determined 
solely from the acceleration parameter $\Gamma_0 $. Lastly, the mathematical expression 
of the emission is the black-body radiation with the temperature $k_B T_U=\hbar \Gamma_0/2\pi$ 
in the upper half volume and is completely different from thermal radiation in the lower half volume. 
We note a similar result was obtained and discussed in Ref. \cite{ADM}. 

When the emission is restricted only for the low momenta, i.e., for the linear 
dispersion regime $\omega_k\simeq kc$ and $kc\ll \Gamma_0 $, the above two cases show 
the same asymptotic behavior. The number of emitted bogolons between $\v k$ and $\v{k}+d\v{k}$ 
in the thermodynamic volume limit is given by 
\be \label{asymp1}
d\tilde{n}_{\v k}\simeq\frac{\lambda^2}{2g\hbar c}\; k\; dk d\Omega \quad {\rm for}\ k\ll \xi^{-1}, \Gamma_0/c,
\ee
where $d\Omega$ is the element of solid angle. 
Although this formula resembles the Rayleigh-Einstein-Jeans law for the black-body radiation, it cannot be 
interpreted as a signature of the thermal radiation. The reason is simply because the effective 
``temperature'' $T_U\propto \Gamma_0 $ does not show up for the long wavelength approximation. 

A serious difficulty arises for the ultraviolet region where the emitted bogolons 
within an infinitesimal wave number $d\v{k}$ is 
\begin{equation}\label{asymp2}
d\tilde{n}_{\v k} \simeq\frac{4\pi gn^{2}\lambda^{2}}{\hbar^{3}\Gamma_{0}c^{2}} dk d\Omega\times
\begin{cases}
e^{-2\pi \omega_k /\Gamma_0}\ ( k_z \zeta_0>0 )\\
1\!+\!e^{-2\pi \omega_k /\Gamma_0}\ (  k_z \zeta_0<0)
\end{cases},
\end{equation}
for $k \gg\xi^{-1}, \Gamma_0/c$.
This becomes a constant and is independent of the excitation energy of bogolons for the lower half volume 
as $k$ approaches infinity.  Therefore, the ultraviolet divergent for the total dissipated energy occurs 
for this half volume when integrating all over the wave numbers. 

One way to understand the origin of the above mentioned ultraviolet divergence is to look at 
the form of potential corresponding to the trajectory \eqref{ther}. 
With a simple analysis of classical mechanics, the potential is 
$V(\zeta)=-M_{imp}\Gamma_{0}^{2}\zeta^{2}/2$ with $M_{imp}$ the mass of impurity, 
which is an inverted harmonic potential. It is clear that the speed of impurity is not bounded 
and can become infinite for the asymptotic past and/or future. In reality, it is impossible 
to construct such a potential for the entire space since it requires an infinite amount of energy. 
Thus, this type of trajectories is only possible within a finite region and hence is 
limited by a given experimental situation, and one should not integrate the whole trajectory 
in the integral \eqref{int}. 

For a finite integral interval specified with $(t_i,\,t_f)$, the integral of interest is calculated 
by the same regularization procedure as 
\begin{multline} \label{int1-3}
I_{\v k}(t_i,\,t_f)=\frac{\left|k_z\zeta_0\right|^{i \omega_k /\Gamma_0}}{\Gamma_0}
e^{-{\rm sgn} (k_z)\; \pi\omega_k /2 \Gamma_0}\\
\left[\gamma (-i\frac{\omega_k }{\Gamma_0},i k_z\zeta_0e^{-\Gamma_0 t_i})
-\gamma (-i\frac{\omega_k }{\Gamma_0},i k_z\zeta_0e^{-\Gamma_0 t_f}) \right],  
\end{multline}
where $\gamma(x,z)$ is the incomplete gamma function of the first kind. 
We note that in the limit $t_i\to-\infty$, the first term in \eqref{int1-3} converges to 
the previous result \eqref{int1-2}. However, the second term oscillates with the angular frequency 
$\omega_k$ as the final time $t_f$ approaches $\infty$. This behavior indicates that the regularization 
and infinite time integral limit do not commute each other and physics for finite time interval may be 
different from the above idealized infinite time interval limit. 
To analyze further, we consider a special case $t_i=0$ and set $t_f=T$. 
Using the asymptotic expansions of the incomplete gamma function, the number of emitted 
bogolons takes in the long wavelength regime as 
\be
d\tilde{n}_{\v k}\simeq\frac{\lambda^2 cT^2}{2g\hbar}\; k^3\; dk d\Omega \quad {\rm for}\ k\ll \xi^{-1}, \Gamma_0/c,
\ee 
whereas it behaves as in the short wavelength regime 
\be
d\tilde{n}_{\v k} \simeq\frac{4 n^{3}g^2\lambda^{2}}{\hbar^{4}c^4}\left[1\!-\!\cos k_z\zeta_0(1-e^{-\Gamma_0 T}) \right]
k^{-2}dk d\Omega,
\ee
for $k \gg\xi^{-1}, \Gamma_0/c$.

Comparing with the previous expressions (\ref{asymp1}, \ref{asymp2}), the finite interval 
case is totally different from the infinite case both for the long and the short wavelength regimes. 
Importantly, $d\tilde{n}_{\v k}$ does not decay exponentially for large value of wave numbers but 
decays algebraically. Therefore, the trajectory \eqref{ther} cannot be used to observe 
the temperature-like radiation unless its acceleration covers an infinite time period.   
This is, however, impossible in real experiments and we conclude that 
the appearance of mathematical expression for the Planckian 
distribution \eqref{result1} is an artifact of invalid extrapolation for the physical parameters. 
%We emphasize that a mathematically valid procedure may not neccesary lead to the correct physical result. 

%%%%%%%%%%%%%%%%%%%%%%%%%%%%%%%
\subsection{Relativistic uniform acceleration}
%%%%%%%%%%%%%%%%%%%%%%%%%%%%%%%

The relativistic trajectory of a uniformly accelerated particle along the $z$-direction is
\be \label{linear}
\v \zeta (t)=\frac{c^2}{a}\sqrt{1+(\frac{at}{c})^2}\;(0,0,1),
\ee
with a given proper acceleration $a$. To discuss a simulation of the relativistic effect in BECs, 
we replace the speed of light by the speed of sound $c$ in the condensate. 
Note that this trajectory can be realized with a non-relativistic potential of the form 
$V(\zeta)=\frac 12 M_{imp}(c^{3}/a\zeta)^{2}$. Importantly, the speed of the impurity is always bounded 
by that of sound, $|\dot{\zeta}(t)|<c$. In the limit $c\to\infty$ or within the short time period $at/c\ll1$, 
the trajectory becomes a non-relativistic quadratic one such as $at^2/2+$(constant).
To evaluate the above integral (\ref{int}), we change the integration variable $t$ to a dimensionless
variable $s$ through $t=(c/a)\sinh s$, which is called a proper time in relativity, and we define
\begin{align}\label{mu}
\mu_{\v k} &=\frac ca \sqrt{\omega_k ^2- (c k_z)^2}=\frac{kc^2}{a}\sqrt{\sin^2 \theta+ k^2\xi^2},\\
\sigma_{\v k} &= \tanh ^{-1} (\frac{c k_z}{\omega_k}),\label{sigma}
\end{align}
where $\theta$ is an angle between $\v k$ and the $z$ axis. Then,

\be
\label{intacc}
I_{\v k}= \frac{2c}{a} [ \pi\;\delta (\mu_{\v k}) \cosh \sigma_{\v k} + i {\rm K}_1 (\mu_{\v k}) \sinh \sigma_{\v k}  ],
\ee
where ${\rm K}_j (x)$ is the $j$th modified Bessel function of the second kind. 
The first term in (\ref{intacc}) will not contribute to the occupation number for 
the following reasons. The first is that $\mu_{\v k}=0$ if and only if $\v k=0$. 
The second reason follows from two properties of the 
Dirac delta distribution; $\delta(f(x))=\sum_i \delta(x-x_i)/|f'(x_i)|$, where 
$x_i$ are the roots of $f(x)=0$, and $x\delta(x)=0$. 
The final reason is the presence of the factor $\epsilon_k/\omega_k$ in (\ref{occu}).
Therefore the occupation number in the thermodynamic limit is  
\be \label{dn1}
d \tilde{n}_{\v k}= \frac{2 n \lambda ^2 c^6k^6\cos^2\theta}{(2\pi)^3Ma^4\hbar\omega_k}
\left(\frac{{\rm K}_1 (\mu_{\v k})}{\mu _{\v k}}\right)^2 dkd\Omega. 
\ee

Let us now examine the result for the infrared and the ultraviolet region 
using the asymptotic behaviors of ${\rm K}_j (x)$. The $\mu_{\v k}$ becomes very small 
in the infrared region, and hence the leading term of (\ref{dn1}) becomes 
\be \label{dn2}
d \tilde{n}_{\v k}\simeq \frac{2 n \lambda ^2 k\cos^2\theta}
{(2\pi)^3Mc^3\hbar(\sin^2 \theta+ k^2\xi^2)^2} dkd\Omega.  
\ee
Therefore, the leading term has a singularity around $k\simeq 0$ at the angles $\theta=0,\pi$, 
i.e., along the direction of trajectory, which results in the divergent result for the total number of emitted bogolons. 
This infrared singularity due to bogolons has the same origin as the famous 
infrared catastrophe in quantum electrodynamics \cite{itzykson}. There, the total number 
of emitted photons due to a motion of accelerated charge particle diverges as $k\to 0$. However, 
the physical measured quantity, the radiated energy, is still finite. 

We next use the asymptotic form of ${\rm K}_1$ for large $\mu_{\v k}$ to get the leading term: 
\be \label{dn3}
d \tilde{n}_{\v k} \simeq \frac{n\lambda^2(2Mc)^3\cos^2\theta}{(2\pi)^2\hbar^5ak^2}\exp(-\frac{\hbar k^2c}{Ma}) dkd\Omega.
\ee
Hence the dissipated energy in the ultraviolet region after integrating over the range of solid angles is 
\be \label{dissacc}
d{\cal E}_{k} \simeq \frac{4n\lambda^2M^2c^3}{ 3\pi \hbar^3a}\exp(-\frac{\hbar k^2c}{Ma}) dk.
\ee
Thus $d{\cal E}_{{\rm tot}}$ asymptotically depends on $k$ through the exponential factor $\exp [-\hbar k^2c/(Ma) ]$. 
We note that this factor can also be obtained by evaluating the integral using the stationary 
phase method. This high momentum behavior could be used to measure the number $\hbar c/M$. 

Lastly we notice that two length scales in this example have different orders of magnitudes; 
the coherence length $\xi$ and the acceleration length $l_a=c^2/a$. 
In most real experimental situations $\xi \ll l_a$, since accelerations available in laboratories are very restricted. 
Then we again use the asymptotic form of ${\rm K}_1$ to estimate the total dissipated energy 
for the weak acceleration limit as 
\be
{\cal E}_{{\rm tot}} \simeq \frac{ n \lambda ^2 M^2 c}{10 \hbar ^3} \sqrt{\frac{\hbar a}{\pi M c^3}} .
\ee
Although the speed of impurities never exceed the speed of sound we expect 
a small finite amount of radiation due to the uniformly accelerated impurity.

\section{Discussion}
%%%%%%%%%%%%%%%%%%%%%%%%%%%%%%%

In this paper, we have investigated a simple model for the accelerated motions of classical impurity 
in the homogeneous condensate and radiations accompanied by them. 
It is shown that the impurity couples linearly to the field of the Bogoliubov excitations 
as shown in \eqref{hb}. Since the effective theory of the low energy excitations are known to resemble 
a relativistic mass-less field theory \cite{analog}, one naively expects that this model can be used to 
simulate the FDU radiation for a suitably chosen accelerated motion of impurity. Indeed, our model 
also shares similar mathematical formulas with other proposed models as discussed 
in Refs.~\cite{SKBFC,ADM,RCPR,MBM,AM}. 

We have, however, pointed out that the appearance of the Planckian distribution for 
a radiation spectra corresponding to the impurity trajectory \eqref{ther} is valid only 
in the mathematical limit in which the impurity is accelerated from the infinitely past to the infinitely remote future. 
This limit requires infinite amount of energy and hence 
one can hardly expect it to be simulated in laboratories. To check whether or not 
any trace of such an extreme situation can be observed in real experiment, 
we have examined the finite acceleration time case 
in detail and have shown that the results are completely different from the idealized limit. 
Obtaining the Planckian distribution also requires an appropriate regularization, 
otherwise the integral does not converge to the expression \eqref{int1-2}. This regularization 
can be understood, for example, as the adiabatic switching off of the impurity coupling, 
i.e., $\lambda\to\lambda \exp(-\epsilon t)$ with $\epsilon\to0$ at the end of calculation. 
We, however, emphasize that the usage of other adiabatic factors will result in 
different formulae and hence the appearance of the Planckian distribution 
is regularization dependent. 
Our result indicates that similar difficulties may exist for observing 
the Planckian distribution in other mentioned proposals. 

We have also examined a relativistic trajectory for a uniformly accelerated impurity 
and have analyzed the radiation spectrum in detail. The result \eqref{dissacc} shows an exponential decay 
of the emitted excitations for short wavelength behavior of the radiation spectrum. 
This exponential factor is sometimes referred to as a Boltzmann factor of the Bose-Einstein distribution 
and is related to the effective temperature of the FDU effect. With this 
correspondence, the effective temperature here is $k_B T= \hbar a/2c$ which differs by a factor 
$\pi$ from the original FDU temperature $T_U=\hbar \Gamma_0/2\pi$. 
It is questionable if we can relate an appearance of the Boltzmann factor to the black-body radiation 
which is an incoherent radiation from the thermal equilibrium state. 
This is because radiations from the motions of impurity in condensate are coherent processes. 
The same statement also holds for the FDU effect. The thermal nature of reservoir 
is a consequence of a statistical ensemble average. From this averaging procedure, the expectation value 
of occupation number for bosons results in the Planckian distribution, but not the other way around.  
In other words, an apparent mathematical expression {\it does not} mean that the system is 
in a thermal equilibrium at certain temperature. This point was originally discussed in Ref. \cite{lee}. 

Along the same line of analysis, we gave a preliminary study for a radiation spectrum 
from a constant circular motion \cite{jun2,jun3}. 
It is expected that this kind of impurity motion can be realized in the current experimental techniques. 
In future, we intend to study the circular motion in more detail based on realistic settings such as 
including a trapping potential, finite temperature effects, the back reaction to the impurity, and so on. 
It is also interesting to examine a connection to the so-called the circular FDU radiation 
and to discuss a possibility to simulate such an effect in BECs. 

Finally, we briefly comment on a possible connection to the emergent theories of gravity. 
Several authors have proposed that gravitation is considered as an emergent phenomenon 
arising from collective excitations in quantum many-body systems 
\cite{mazur0,mazur1,mazur2,mazur3,mazur4,mazur5,mazur6,mazur7,chapline,laughlin}. 
In these proposals a physical vacuum is considered as a real condensed state 
which contains a huge number of heavy bosons, and the Minkowski vacuum is assumed 
to be the homogeneous condensate. It is clear that effective theory breaks down 
at a certain energy scale and the deviations from the prediction of 
the relativistic quantum field theory are expected to occur. In this line of research, one is interested in 
these deviations and the new phenomena predicted by the microscopic theories behind, 
that are possibly observed in astrophysics. 

We are then interested in looking at the radiation for beyond the linear dispersion regime. 
In the relativistic uniform acceleration, for example, we see that the short wavelength behavior 
of the radiation spectrum is proportional to $\exp(-\mu_k)\simeq\exp(-\omega_k c/a)$ along the direction of the trajectory. 
This dependence can be used to estimate higher modes of the excitation spectrum 
of the Bogoliubov excitations. It is also an open question to investigate if one can simulate 
the FDU effect itself in BECs, in which {\it any accelerated observer} perceives 
the Minkowski vacuum as a thermal state, and to clarify the physical mechanism 
leading to this effect if it exists. 

%%%%%%%%%%%%%%%%%%%%%%%%%%%%%%%%%%%%%%
\begin{acknowledgements}
The author would like to express his acknowledgment to Professor Pawel O. Mazur for
suggesting this problem, many valuable discussions, 
explaining the physical basis of Refs. \cite{mazur0, mazur1, mazur2, mazur3, mazur4}, 
and providing additional information about these references. 
This work was partially supported by MEXT. 
\end{acknowledgements}
%\section*{References}
%%%%%%%%%%%%%%%%%%%%%%%%%%%%%%%%%%%%%%

%%%%%%%%%%%%%%%%%%%%%%%%%%%%%%%%%%%%%%


\begin{thebibliography}{10}
\bibitem{NCmethod} There are several proposals for the number conserving treatment for BEC 
\cite{girardeau, gardiner, castin}.  We follow the one used in refs.~\cite{girardeau}.
\end{thebibliography}
\end{document}